\documentclass{elsart}
\usepackage{graphics}
\newcommand{\g}{\gamma}

\newcommand{\da}{\dagger}  % symbol for Hermitian conjugate dagger
\newcommand{\be}{\begin{equation}}
\newcommand{\eq}{\end{equation}}
   
\newcommand{\Tr}{{\rm \, Tr \!}}    %symbol for trace
\begin{document}
\noindent DAMTP-2000-67 
\begin{frontmatter}
\hyphenation{Coul-omb ei-gen-val-ue ei-gen-func-tion Ha-mil-to-ni-an
  trans-ver-sal mo-men-tum re-nor-ma-li-zed mas-ses sym-me-tri-za-tion
  dis-cre-ti-za-tion dia-go-na-li-za-tion in-ter-val pro-ba-bi-li-ty
  ha-dro-nic he-li-ci-ty Yu-ka-wa con-si-de-ra-tions spec-tra
  spec-trum cor-res-pond-ing-ly}
\title{Transverse lattice}\footnote{Invited talk at Xth 
International Light-Cone Meeting on Non-Perturbative QCD 
and Hadron Phenomenology, Heidelberg 12-17 June 2000. To appear in
proceedings.}
\author{S. Dalley}\footnote{supported by PPARC grant No.\ GR/LO3965.}
\address{Centre for Mathematical Sciences, \\
Wilberforce Road, Cambridge CB3 0WA, England \\ sd214@damtp.cam.ac.uk}

\date{30 June 2000}
\begin{abstract}
The transverse lattice approach to non-perturbative light-front
hamiltonian QCD is described. Preliminary  results on the $\pi-\rho$
system are presented, at fixed DLCQ and Tamm-Dancoff cut-offs. 
A renormalised, approximately
Lorentz covariant light-front hamiltonian is found to 
leading order of the colour-dielectric expansion, compatible
with a massless pion. 
The $\pi$ light-front wavefunction is
compared with experiment. Exclusive processes agree reasonably well, 
given the approximations, but
inclusive processes, sensitive to higher Fock state structure, 
still exhibit large cut-off artifacts.  
\end{abstract}
\end{frontmatter}

%%%%%%%%%%%%%%%%%%%%%%%%%%%%%%%%%%%%%%%%%%%%%%%%%%%%%%%%%%%%%%%%%%%%%%
%%%%%%%%%%%%%%%%%%%%%%%%%%%%%%%%%%%%%%%%%%%%%%%%%%%%%%%%%%%%%%%%%%%%%%
%%%%%%%%%%%%%%%%%%%%%%%%%%%%%%%%%%%%%%%%%%%%%%%%%%%%%%%%%%%%%%%%%%%%%%

\section{Introduction}
\label{intro}

A light-front hamiltonian has wavefunctions initialised on a null
plane $x^+ = (x^0 + x^3)/\sqrt{2} = 0$.
They are manifestly Lorentz-boost invariant and therefore
suitable for highly relativistic problems. Indeed, 
many high energy QCD scattering or decay processes factorize into a 
non-perturbative 
light-front wavefunction convoluted with a perturbative scattering
kernel \cite{stan2}.
The calculation of light-front wavefunctions
in QCD is not easy, however. One has to address the usual
problems of confinement and chiral symmetry breaking in an
unusual theoretical context, where renormalisation issues are
non-standard and unavoidably non-perturbative.  

Path integral lattice gauge theory simulation has been the most comprehensive
method to compute hadronic observables from first principles QCD. 
Some indirect information on
light-front wavefunctions has also been obtained from this method, but
far less than one would like. It is therefore natural to attempt a 
light-front hamiltonian quantisation of lattice gauge theory. 
Treating $x^+$ as
canonical time and $\{x^-= (x^0 - x^3)/\sqrt{2}, x^1, x^2 \}$ as
the spatial variables, $x^+$ must be continuous for a light-cone
hamiltonian formulation. A high-energy lattice cut-off can only reasonably
be applied to the transverse spatial directions ${\bf x} = \{x^1, x^2\}$,
since large values of the momentum $p^+$ conjugate to $x^-$ correspond
to small energies $p^-$ conjugate to $x^+$. 
A high energy cut-off may be applied to
$x^-$ by making it periodic, $x^- \equiv x^- + L$. For a state of
total longitudinal momentum $P^+$, the choice $L = 2 \pi K / P^+$,
with $K$ a fixed integer, is sometimes convenient (discrete
light-cone quantisation (DLCQ) \cite{dlcq}). 
Naively, one might try to 
obtain the corresponding transverse lattice gauge theory by
taking the small distance continuum
limit of Wilson's lattice gauge theory \cite{wilson2} in the $x^+$
and $x^-$ directions. However, there are some difficulties associated
with this; in particular the light-cone quantisation of the resulting
coupled gauged non-linear sigma models at each transverse lattice
site is awkward \cite{grif}. 
Bardeen and Pearson \cite{bard} suggested that, for
large values of the transverse lattice spacing $a$, a linear
sigma model approximation could be used. Instead of formulating the
lattice part of the theory in terms of link variables $U \in SU(N)$, one uses
link variables $M \in GL(N)$, where the complex 
matrices $M$ transform in the same way under gauge transformation.
Use of such linearized variables is expected to be efficient on
coarse lattices, as they represent the important degrees of freedom
at such scales. Further discussion of the origin and meaning
of these `colour-dielectric' variables can be traced in ref.~\cite{hans}.

Thus, in transverse lattice gauge theory the Lorentz indices 
$\mu, \nu \in \{ 0,1,2,3 \}$ are
split into LF indices $\alpha,\beta \in \{+,-\}$
and transverse indices $r,s\in \{1,2\}$. One has link variables
$M_r({\bf x})$  associated
with the link from ${\bf x}$ to ${\bf x} + a \hat{\bf r}$ on a square
transverse lattice, together
with continuum $SU(N)$ gauge potentials $A_{\alpha}({\bf x})$ and
Dirac fermions $\Psi({\bf x})$ 
associated with a site ${\bf x}$.
These variables transform under transverse lattice gauge
transformations $V \in SU(N)$ as
\begin{eqnarray}
        A_{\alpha}({\bf x}) & \to & V({\bf x}) A_{\alpha}({\bf x}) 
        V^{\da}({\bf x}) + {\rm i} \left(\partial_{\alpha} V({\bf x})\right) 
        V^{\da}({\bf x})  \nonumber \\
        M_r({\bf x}) &  \to & V({\bf x}) M_r({\bf x})  
        V^{\da}({\bf x} + a\hat{\bf r}) \\
        \Psi({\bf x}) & \to & V({\bf x})\Psi({\bf x}) \ \label{sym}.
\end{eqnarray}
While a lattice gauge theory at large $a$ 
may have simple solutions when formulated
in terms of disordered variables $M$, this will cease to be true when
$a$ is made small. 
Although one may only be interesting in solving the theory at large
$a$ because of this, traditionally the only way to derive
the appropriate effective theory was to start at small $a$ and
`integrate out' short distance degrees of freedom. This problem
unfortunately remains largely unsolved in the present case. 
A more radical approach
attempts to remove cut-off artifacts already at large $a$, by
tuning the effective theory to respect the continuum symmetries
(gauge, Poincar\'e, chiral) required of the theory. This is the method
that will be followed here. 

The hamiltonian will consist of all operators invariant
under lattice gauge symmetries (\ref{sym}) and Poincar\'e symmetries unviolated
by the cut-offs. We will always have in mind that the transverse
lattice
spacing $a$ is the only cut-off to which this applies, all others,
such as DLCQ, having been extrapolated (in principle). As a result,
we may also use dimensional counting in the $x^{\pm}$ co-ordinates to limit the
number of allowed operators.
We search for a domain of coupling
constants\footnote{We must allow these couplings to be functions of 
longitudinal momentum $p^+$ in general \cite{wilson1}.} 
where  Poincar\'e symmetries violated
by the cut-off $a$ are restored.
In a general lattice gauge theory, gauge and Poincar\'e symmetry 
alone is not sufficient to yield 
a trajectory of couplings flowing to the continuum QCD.
However, the transverse lattice gauge theory is actually
`half-continuum'
($A^+$ and $A^-$ appear explicitly), so it is inconceivable
that anything else is obtained provided Lorentz invariance can be
convincingly demonstrated. Symmetries that may be `spontaneously
broken', such as chiral symmetry, are treated in an unconventional 
way in light-front quantization. Because cut-offs may, and usually
are chosen to, render the vacuum state trivial, symmetries that
would conventionally be referred to as spontaneously broken will
appear explicitly broken in the light-front hamiltonian. The
behaviour of the explicit symmetry-breaking couplings is
governed by the underlying symmetry invariance \cite{wilson1} 
and, in the case
of `dynamical symmetry breaking', by stability
of the vacuum.

In a real calculation at a given lattice spacing $a$ one must make
further 
approximations, both on the
Hamiltonian and its Hilbert (Fock) space. The Fock space at fixed
lattice spacing may be truncated with the
DLCQ cut-off $K$ and also a Tamm-Dancoff cut-off on the 
maximum number of quanta in a state. In principle the latter
two cut-offs should be extrapolated, as has been done in 
studies of glueballs \cite{dvds}, if
we are to follow the line of argument above.
The QCD hamiltonian may be expanded in gauge-invariant powers
of $M$ and the quark fields $\Psi$. The intuitive justification
for this is that both should behave like massive degrees of freedom
(at large $a$); the
former because of its association with the colour-dielectric 
mechanism of confinement and the latter because of spontaneous chiral  symmetry
breaking. We may also assume some transverse locality by expanding
the hamiltonian in $a {\bf P}$, where ${\bf P}$ is transverse
momentum. 
All these approximations have a physical justification
and may be systematically relaxed.
A final couple of approximations, which may be made for
convenience, are 
to keep as many  Poincar\'e generators as possible in kinematic
form (independent of interactions and their renormalisation)
and to neglect non-trivial
 longitudinal momentum dependence of coupling `constants'.
The consequences of this will show up in the extent to which 
symmetry can be restored.
The above approximations provide a general prescription for
reducing the allowed operators in the hamiltonian down to a finite
set, which may be systematically enlarged,
and reducing the Hilbert space to a finite dimension, which may be
extrapolated.
The reward for a non-standard quantization with non-standard
variables and approximations will be a dramatically simplified
hadronic wavefunction.

\section{Light meson calculations}

Let us consider the first non-trivial approximation to the problem
of meson boundstates in transverse lattice QCD.
Applying the 
previous considerations, the leading-order lagrangian for a single
quark flavour is 
\begin{eqnarray}
L & = &  \sum_{{\bf x}} \int dx^- \sum_{\alpha, \beta = +,-}
\sum_{r=1,2} 
-{1 \over 2 G^2} \Tr \left\{ F^{\alpha \beta} F_{\alpha \beta} \right\}
 \nonumber
\\
&& + \Tr\left\{[\left(\partial_{\alpha} +i A_{\alpha} ({\bf x})\right)
        M_r({\bf x})-  i M_r({\bf x})   A_{\alpha}({{\bf x}+a
 \hat{\bf r}})][{\rm h.c.}]\right\}
\nonumber \\
&& - \mu^2  \Tr\left\{M_r M_r^{\da}\right\}
 + i \overline{\Psi} 
\g^{\alpha} (\partial_{\alpha} + i A_{\alpha}) \Psi - \mu_F
\overline{\Psi}\Psi 
\nonumber\\
&& +  i \kappa_A \left( \overline{\Psi}({\bf x}) \g^{r} M_r({\bf x})
 \Psi({\bf x} + a \hat{\bf r}) 
- \overline{\Psi}({\bf x}) \g^{r} M_{r}^{\da}({\bf x}- a \hat{\bf r}) 
\Psi({\bf x} - a \hat{\bf r})
\right)\nonumber\\
&&
+ \kappa_S \left( \overline{\Psi}({\bf x}) M_r({\bf x})
 \Psi({\bf x} + a \hat{\bf r})+\overline{\Psi}({\bf x}) M_{r}^{\da}({\bf
x}- 
a \hat{\bf r})
 \Psi({\bf x} - a \hat{\bf r})\right)\nonumber
\end{eqnarray}
The Hilbert space of the corresponding light-front hamiltonian will
also be expanded in fields. For the present calculation we keep only
the  $|\overline{\Psi}({\bf x})
{\Psi}({\bf x}) >$
and $|\overline{\Psi}({\bf x}) M_r ({\bf x}) \Psi({\bf x} + a \hat{\bf r}) >$ 
states and their translates. 
Thus, in this `one-link' approximation, a quark and antiquark can be
at the same transverse lattice site, or separated by one
link.\footnote{This approximation to the Hamiltonian and Fock space
was first considered in ref.\cite{mat3}. Those authors
subsequently treated it  as a phenomenological model, fixing coupling constants
by hand and comparison to experimental mass ratios. However, see also
Seal's talk in these proceedings.} This is the most
severe approximation that still allows a meson to propagate on the
transverse lattice. 
The calculations I have done so far have been for fixed $K$. This cut-off (at
least) must eventually be extrapolated.

Using the chiral representation \cite{IZ} we decompose 
$\Psi^{\dagger} = (u_{+}^{*}, v_{+}^{*}, v_{-}^{*},
u_{-}^{*})/2^{1/4}$ into left (right) movers $v$ ($u$) with a helicity
subscript.
In light-cone gauge $A_- = 0$,
one can  eliminate non-dynamical degrees of freedom $A_{+}$ and
$v_{\pm}$ at the classical level,
to derive a light-front hamiltonian in terms of transverse
polarizations
only:
\begin{eqnarray}
 P^-  &  = &  \int dx^- \sum_{{\bf x}} 
   {G^2 \over 4} \left(\Tr\left\{ 
              J^{+} \frac{1}{({\rm i} \partial_{-})^{2}} J^{+} \right\}
            -{1 \over N} 
        \Tr\left\{ J^+  \right\} {1 \over ({\rm i}\partial_{-})^{2} }
     \Tr\left\{ J^+ \right\} \right) \nonumber \\
&& + {\mu_{F}^{2} \over 2} \left( F_{+}^{\da} {1 \over {\rm i} \partial_-}
F_+ + F_{-}^{\da} {1 \over {\rm i} \partial_-} F_- \right) 
 + \mu^2  \sum_{r=1}^{2}  \Tr\left\{M_r M_r^{\da}\right\}
\end{eqnarray}
\begin{eqnarray}
F_{\pm}({\bf x}) & = & - u_{\pm}({\bf x}) + {\kappa_S \over \mu_F}
\sum_{r} \left( M_r({\bf x} - a \hat{\bf r}) + M_{r}^{\da}({\bf x} )
  \right)u_{\pm}({\bf x}) \nonumber \\ 
&& \pm 
{{\rm i} \kappa_A \over \sqrt{2} \mu_F} \left((M_{1}({\bf x} - a \hat{\bf r}) 
\mp iM_{2}({\bf x} - a \hat{\bf r})) \right. \nonumber \\
&&   \left. - (M_{1}^{\da}({\bf x} ) \pm iM_{2}^{\da}({\bf x} )) 
\right)u_{\mp}({\bf x})  \\
J^{+}({\bf x}) &=&  i \sum_{r} \left(
M_r ({\bf x}) \stackrel{\leftrightarrow}{\partial}_{-} 
M_r^{\da}({\bf x})  + M_r^{\da}({\bf x} - a\hat{\bf r}) 
\stackrel{\leftrightarrow}{\partial}_{-} M_r({\bf x} - a\hat{\bf r})
\right)  \nonumber \\
&& + u_{+}({\bf x})u_{+}^{\dagger}({\bf x}) + u_{-}({\bf x})
u_{-}^{\dagger}({\bf x})
\end{eqnarray}
With the exception of DLCQ self-energies, which are logarithmically
divergent as $K \to \infty$ and needed for finite physical 
answers, the quartic terms in $F^{\da} ({\rm i} \partial_{-})^{-1} F$
are dropped for the following calculation since the one-link
approximation treats them asymmetrically \cite{mat3}. They may be included in
higher link approximations and are essential for 
recovering parity invariance \cite{mat2}. 
Let us define $\bar{G}=G\sqrt{(N^2-1)/N}$, which has the dimensions of
mass, and introduce the 
dimensionless variables
\be
mb = {\mu \over \bar{G}}\ \ ; \ \
mfs1=\left( {\mu_{F}^{(1)} \over \bar{G} } \right)^2 \ \ ; \ \
mf2 = {\mu_{F}^{(2)} \over \bar{G}} \ ;  \label{coup1}
\eq
\be
ka  = {\kappa_A \over \bar{G}}\ \ ; \ \
ks  = {\kappa_S \over \bar{G}}\ . \label{coup2}
\eq
A Fock-sector dependent bare fermion mass
$\mu_{F}^{(i)}$ has been allowed because the one-link approximation
produces
sector-dependent self-energies.
$i=1$ is the $\bar{\Psi}\Psi$ sector
and $i=2$ corresponds to the $\bar{\Psi}M\Psi$ sector. All
$N$-dependence of the theory has been absorbed into $\bar{G}$ now. 

Following
a successful series of
glueball studies with van de Sande \cite{dvds,lect}, I
investigated the coupling constant space (\ref{coup1})(\ref{coup2}) for signals
of enhanced Lorentz covariance using similar methods. The preliminary results I
present here are mainly intended as an illustration of the procedure,
and in no way should be considered definitive.
Ideally, after DLCQ and Tamm-Dancoff cut-offs are extrapolated,
the theory should exhibit enhanced Lorentz covariance on an
approximately one-dimensional trajectory of couplings, along which $a$ varies.
This trajectory represents the best approximation to the 
(unique) renormalised trajectory in the infinite-dimensional
space of couplings. The estimate itself is not unique because it 
depends on the quantification of violations of Lorentz
covariance. 
Before proceeding one must address chiral symmetry. There are two
related issues. Firstly, cut-offs tend to break chiral symmetry in 
light-front quantisation because it is a dynamical symmetry.
Secondly, spontaneous chiral symmetry breaking cannot appear as a
non-trivial vacuum. The first issue means we should allow
chiral-symmetry breaking terms in the cut-off
hamiltonian. In principle they should be
tuned so that the appropriate chiral Ward identities are satisfied.
Spontaneous breaking would then have to appear as explicit breaking in the
hamiltonian (see \cite{wilson1} for an instructive example). 
Since the boundstate spectrum is at our
disposal, a convenient criterion that addresses both issues at once
is 't Hooft's anomaly matching
condition \cite{hoof}. In a Lorentz-covariant theory, 
chiral symmetry is realised in the boundstate spectrum
either as a goldstone boson (massless $\pi$) or massless composite
fermions. Therefore, we could try to find the renormalised
trajectory by optimising both Lorentz covariance and one or other
of these chiral symmetry conditions. An incorrect choice for the
manifestation of chiral symmetry will make Lorentz covariance
difficult to obtain in a theory with a stable vacuum (no tachyons).
\begin{figure}
\begin{center}
\resizebox{1.0\textwidth}{!}{
 \includegraphics{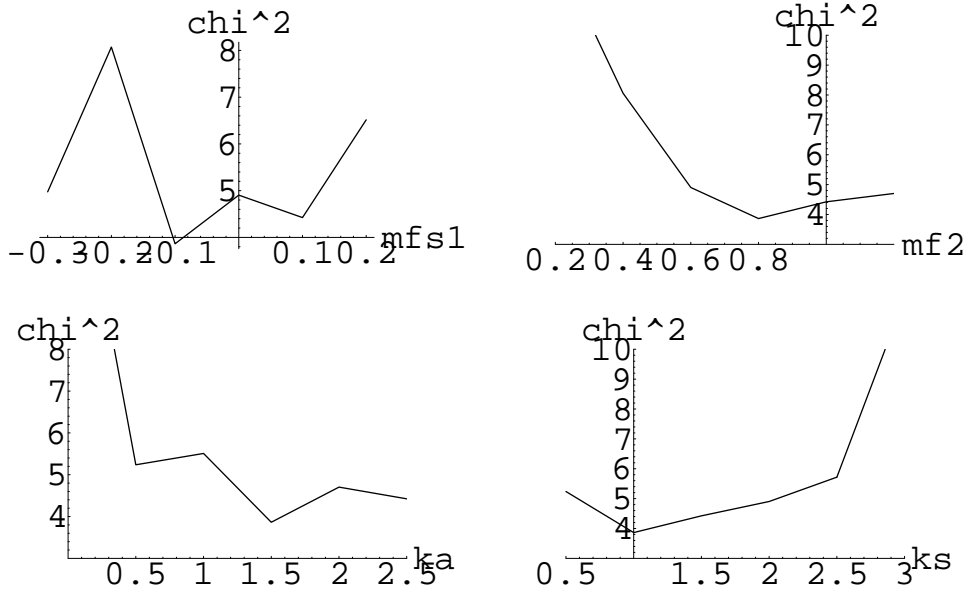}
}\caption{\label{final} 
$\chi^2$ test of Lorentz and chiral symmetry as a function
of couplings for $mb = 0.2$.
   }\end{center}
\end{figure}
I used a $\chi^2$ test with variables to measure
the anisotropy of $\pi$ and $\rho$ dispersion relations and the
deviation from zero of the pion to $\rho$ mass ratio.\footnote{To 
complete the argument
one should also study baryons. This requires more links and fermions
for lattice propagation and chiral anomalies.} 
The QCD scale was set from the (experimental) $\rho$ mass 
and the lattice spacing by demanding isotropy of the $\pi$'s
dispersion in continuum and on-axis lattice directions.
Solving the $P^-$ eigenvalue problem for 
about 1000 combinations of the  couplings (\ref{coup1})(\ref{coup2}) 
and at various 
momenta, figure \ref{final} shows the result for the one-link
approximation at $K=8$ and a particular link-field mass $mb=0.2$. 
The transverse lattice spacing $a=0.35$fm (with a large error). 
This creates a problem in the one-link approximation, since
it is smaller that true $\pi$'s radius. The $\pi$ is artificially
squashed in the transverse direction or, in  more technical language, 
the higher Fock state
structure will be prone to cut-off artifacts. The 
$|\overline{\Psi}({\bf x})
{\Psi}({\bf x}) >$ sector may be less sensitive to cut-off artifacts,
however.

Although the results are still quite crude and further work is needed
to search couplings more finely and with more realistic criteria,
extrapolate $K$ and add more links/quarks,
some results are given here 
for experimentally known observables.
The spin 
projections of the $\rho$ are not all degenerate because of residual
breaking of Lorentz symmetry. Setting the averaged mass to
770 MeV, at the couplings with minimum $\chi^2$
I find $m_{\pi} = 23$ MeV, 
$m_{\rho}(J_z=0) = 642$MeV, $m_{\rho}(J_z=\pm 1) =
898$MeV.
\footnote{As a consistency check on the scale, 
the string tension calculated in the same approximation is found to be
$\sqrt{\sigma} = 0.643 m_{\rho} = 495$MeV.}
The $\pi-\rho$ splitting is generated by the helicity-flip term 
$\kappa_A$ (see Perry's
talk in these proceedings). 
The $|\overline{\Psi}({\bf x}) {\Psi}({\bf x}) >$ component of the
meson wavefunction is  related to the leading order
perturbative QCD expression for many exclusive processes
\cite{stan1}. Fitting the transverse lattice pion wavefunction
to the conformal expansion \cite{conf1,conf2}  of this component, one finds
\begin{eqnarray}
\phi_{\pi}(x, Q^2 \sim 1 GeV^2) & = & {2.653 \over a}
x (1-x) \left\{ C_{0}^{3/2} + 0.237 C_{2}^{3/2} - 0.102 C_{4}^{3/2}
\nonumber \right. \\
&& \left. -0.05 C_{6}^{3/2} + \cdots \right\} \ . \label{exp}
\end{eqnarray}
The transverse scale $1$GeV is a rough estimate based on $\pi/a$,
and $C_{n}^{3/2}(1-2x^2)$ are the appropriate Gegenbauer polynomials.
Eq. (\ref{exp}) directly confirms that the conformal expansion is a good
one. The overall normalisation yields $f_{\pi} = 101$MeV
compared with the experimental value $f_{\pi}(exp.) = 93$MeV. 
\begin{figure}
\begin{center}
\resizebox{1.0\textwidth}{!}{
 \includegraphics{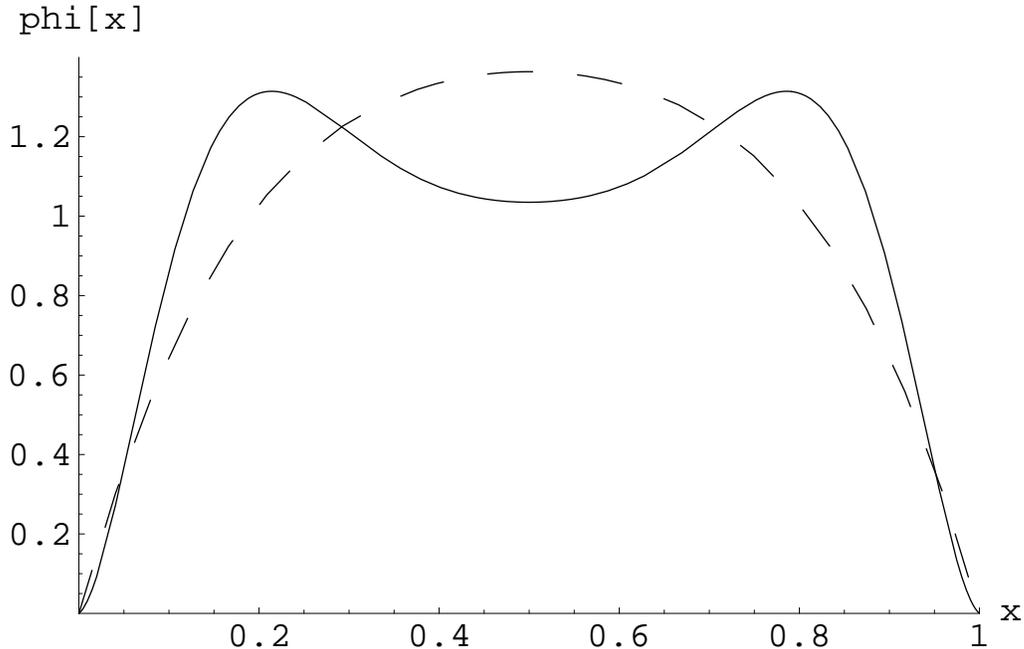}
}\caption{\label{form} 
The pion valence quark amplitude at zero transverse
separation and $Q^2 \sim 10 GeV^2$. 
Solid line is theory, chain line is experimental deduction
(both with unknown error!). Both are normalised to area 1 for
this comparison.
   }\end{center}
\end{figure}
Using leading order perturbative QCD evolution to other scales $Q^2$, and
assuming leading order perturbative QCD factorization, we can compare
(\ref{exp}) with direct and indirect experimental 
measurements. The cleanest extraction from a $\pi$ form factor comes
from $\g \g^* \to \pi^0$; $Q^2 F_{\pi^0}$ is approximately constant
in the range $1 GeV^2  < Q^2 < 10 GeV^2$ and at the higher end
$Q^2 F_{\pi^0} (Q^2 = 8 GeV^2) = 0.16 \pm 0.03$GeV has been
measured at CLEO-II \cite{cleo}. This compares with the theoretical
result 
\be
Q^2 F_{\pi^0} (Q^2 = 8 GeV^2) = {4  \over \sqrt{3}} \int_{0}^{1}
dx {\phi_{\pi}(x, Q^2 \sim 8 GeV^2) \over x} = 0.21{\rm  GeV}  \label{theory}
\eq
Direct tests of $\phi_{\pi}$ have recently become possible
from diffractive dissociation on a nucleus $\pi + A \to A + {\rm
jets}$ \cite{ashery}. Figure \ref{form}
 compares the  quark amplitude with the one 
quoted in ref.\cite{ashery} as that which  best fits the jet data after
hadronization and experimental acceptance (see Ashery's talk in these
proceedings for an updated account of this experiment.) 
Although hadronization tends
to wash out any fine structure in the quark amplitude, making
true comparison between similar curves difficult, the theoretical
curve seems a bit too peaked away from $x=1/2$. This is consistent
with the result (\ref{theory}) being slightly too high.
Of course there is no reason why the finite-$K$ calculation
should give exactly the right result. Use of a 
nearly massless pion, ambiguous normalisation scale $Q^2$, leading
order evolution from 1 to $10 GeV^2$, etc. leads to further errors.
\begin{figure}
\begin{center}
\resizebox{1.0\textwidth}{!}{
 \includegraphics{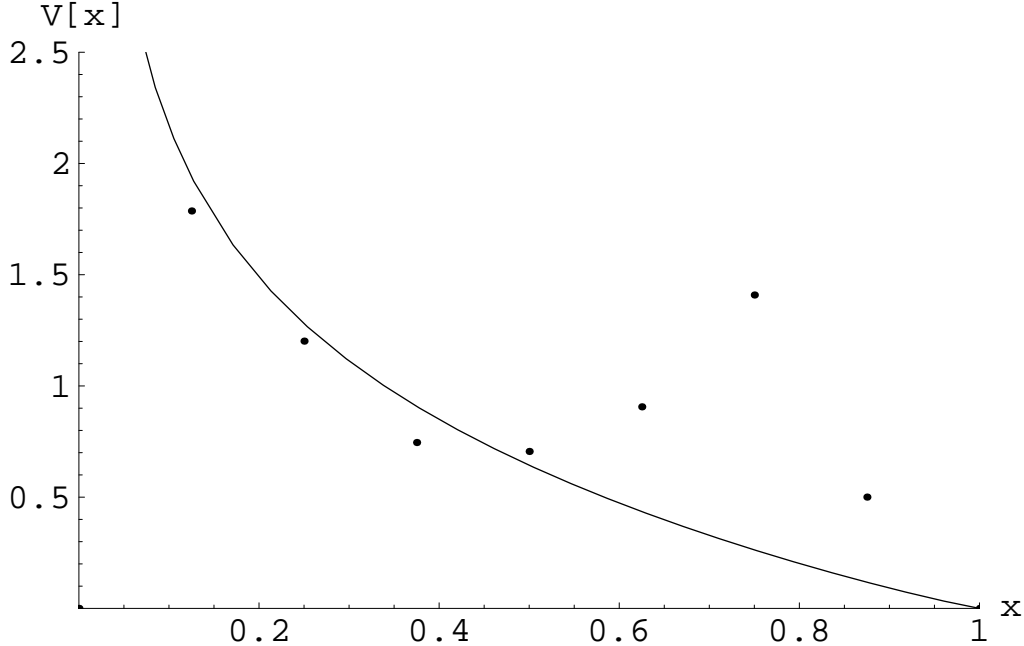}
}\caption{\label{probden}
The non-singlet quark probability distribution $V$ of the pion at
$Q^2 \sim 4 GeV^2$.
Solid line is a model fit to experiment, data points are from the
transverse lattice wavefunction with leading order evolution.
   }\end{center}
\end{figure}

Inclusive processes are typically sensitive to higher Fock state
structure. Here, we may not expect to do so well because of the
severity of our cut-offs. This suspicion is confirmed by the standard
$x^{\alpha}(1-x)^{\beta}$ fits to the non-singlet quark probability
distribution in the $\pi$, obtained from $\pi N$ Drell-Yan \cite{sutton};
see fig.\ref{probden}. The deviation of the lattice results 
at large $x$ is due at least partly to
cut-off artifacts. Relaxing the DLCQ and one-link approximation
will allow some of the quark momentum at large $x$ to 
radiate into small $x$ quarks and gluons as more gluonic
channels are opened. The gluons carry only about $10 \%$ of the $\pi$ momentum
at present --- much lower than the accepted experimental value \cite{na3} ---
confirming there is a problem with higher Fock states.

\end{document}